# Optimization of Coupling between Photonic Crystal Resonator and Curved Microfiber

In-Kag Hwang, Guk-Hyun Kim, and Yong-Hee Lee

*Abstract*— The evanescent coupling from a photonic crystal resonator to a micron-thick optical fiber is investigated in detail by using a 3D-FDTD method. Properly designed photonic crystal cavity and taper structures are proposed, and optimal operating conditions are found to enhance the coupling strength while suppressing other cavity losses including the coupling to the slab propagating mode and to the higher-order fiber mode. In simulation, the coupling into the fundamental fiber mode is discriminated from other cavity losses by spatial and parity filtering of the FDTD results. The coupling efficiency of more than 80% into the fundamental fiber mode together with a total Q factor of 5,200 is achieved for the fiber diameter of 1.0 μm and the air gap of 200 nm between the fiber and the cavity.

## I. Introduction

Recent progresses of photonic crystal technology made it possible to realize ultrasmall, ultrahigh-Q optical cavities which are of interest in quantum optics [1,2], nonlinear optics [3,4], lasers [5], and optical communications [6]. The optical interface of those cavities with external optics is now an important issue for their practical applications. The most common way is to use cavity-to-waveguide coupling followed by waveguide-optical fiber coupling. However, due to the considerable coupling loss coming from the large index difference and the mode size mismatch at the waveguide-fiber junction, very delicate coupling element is required to reduce the loss below a tolerable level. The direct evanescent coupling between the cavity and a tapered optical fiber has been also proposed and demonstrated [7]. The direct optical fiber coupling may be better fitted to many applications especially for a single isolated device because of its negligible propagation loss, direct interconnection with external optical systems, and tunability of the coupling strength. The significant drawback, however, is found in its poor coupling efficiency. It mainly comes from large difference of refractive indices of silica fiber ($n$~1.5) and of semiconductor($n$~3.4)-based photonic crystal cavity, which results in inefficient mode overlap and poor phase matching between the fiber and the cavity modes. Very recently, we reported that the fiber coupling strength can be significantly increased by proper design of the photonic crystal cavity [8], and highly efficient (estimated to ~70%) output collection from the cavity into the fiber was experimentally demonstrated [9]. The coupling scheme is depicted in Fig. 1. A microfiber prepared by tapering conventional single mode fiber down to ~1 μm is positioned above the photonic crystal cavity with an air gap of typically less than 1 μm. The microfiber is sharply bent with a radius of ~ 50 μm to be able to selectively access the photonic crystal cavity of interest without unwanted interaction with other regions. The alignment of the curved feature of the fiber also eliminates the requirement of delicate alignment of the taper.

There are several issues that require special attention when one wishes to achieve optimal fiber coupling for specific applications. In this paper, the detailed coupling mechanism of the cavity-fiber system is investigated based on the finite-difference time domain (FDTD) simulation, and optimal coupling structures and operating conditions are proposed. This study also reveals the limiting factors hindering the achievement of nearly 100% coupling efficiency.

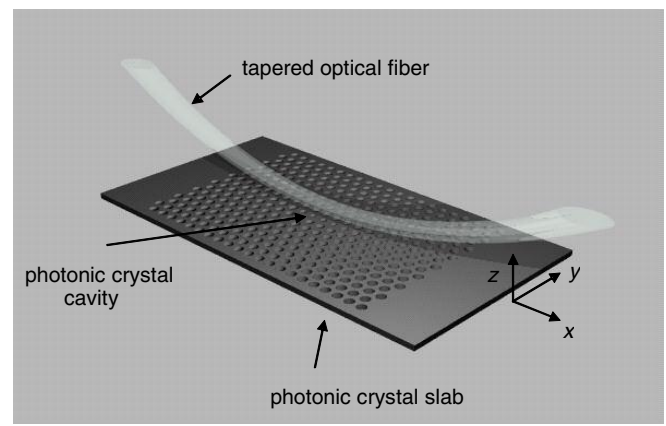

Fig. 1. Schematic diagram of cavity-fiber coupled system. A tapered optical fiber is sharply bent and positioned above a photonic crystal cavity with a sub-micron air gap for the evanescent coupling between the fiber and the cavity modes.

## II. Conditions of Efficient Fiber Coupling

Total quality factor ($Q_{tot}$) of the coupled system of Fig. 1 can be decomposed into several components as follows, depending on the route that photons take to exit the cavity.

Manuscript received August 11, 2005. This work was supported by National R&D Project for Nano Science and Technology, and by Ministry of Science and Technology(MOST) of Korea.
The authors are with the Department of Physiscs, Korea Advanced Institute of Science and Technology, Daejeon, Korea. (e-mail: ikhwang@kaist.ac.kr)

$$\frac{1}{Q_{tot}} = \frac{1}{Q_{air}} + \frac{1}{Q_{slab}} + \frac{1}{Q_{fiber}}$$

$$where \quad \frac{1}{Q_{fiber}} = \frac{1}{Q_{LP01}} + \frac{1}{Q_{LP11}}$$

(1)

Each term is proportional to coupling strength to air ($1/Q_{air}$), slab($1/Q_{slab}$), and fiber($1/Q_{fiber}$), and the fiber coupling is divided again into the couplings to fundamental ($LP_{01}$) and first higher-order ($LP_{11}$) modes of the microfiber. Coupling to other higher-order fiber modes is negligible when the taper diameter is less than a few μm. $Q_{air}$ is not significantly affected by presence of the microfiber, and thus almost fixed at the intrinsic $Q$ of the cavity. Coupling to the slab mode($1/Q_{slab}$) is usually negligibly small because of the photonic bandgap formed along the in-plane direction, but, it dramatically increases as the microfiber approaches the cavity. $Q_{LP01}$ and $Q_{LP11}$ are determined by their mode overlaps with the resonant cavity mode, and dependent on various parameters such as the cavity structure, the fiber diameter and the air gap between the microfiber and the cavity. Since the microfiber converges to a single mode fiber at the end of the taper, any higher-order mode coupled from the cavity will be eventually lost and only the light in the fundamental mode can contribute to the fiber output. Thus, the photon coupling efficiency ($\eta$) from the cavity into the fiber is defined by

$$\eta = \frac{1/Q_{LP01}}{1/Q_{tot}} = \frac{Q_{tot}}{Q_{LP01}}.$$

(2)

To achieve a high $\eta$, the coupling strength for $LP_{01}$ mode should be made much stronger than other couplings, or $Q_{LP01} \ll Q_{air}, Q_{slab}, Q_{LP11}$. It can be realized by proper design of photonic crystal cavity and microfiber structures, and optimization of the coupling conditions. In advance of simulations of the fiber coupling in various conditions, the modal characteristics of the fiber and the cavity are investigated first to get better understanding of the following simulation results.

The effective refractive indices of the $LP_{01}$ and the $LP_{11}$ microfiber modes are shown in Fig. 2 (a) as a function of fiber diameter normalized by wavelength. There are two degenerate modes with orthogonal polarizations in the $LP_{01}$ mode, and four almost-degenerate modes with different polarizations and intensity distributions in the $LP_{11}$ mode[10]. Since the photonic crystal cavity mode of interest has an even $E_y$ field respect to the x-z plane, it couples with the fiber modes that have even $E_y$ component in their evanescent field. This condition is satisfied in only two fiber modes, each one from $LP_{01}$ and $LP_{11}$ mode, and their electrical field profiles are shown in Fig. 2 (b). Here we assumed that the photonic crystal slab is located below the taper. The $LP_{11}$ mode cutoff is $D/\lambda=0.75$, which corresponds to the fiber diameter of $D \approx 1.1$ μm for the wavelength of 1520 nm. The coupling strength between the fiber and the cavity modes is determined by the transverse mode overlap in y-z plane and the phase matching along x-axis between the two modes. In terms of the phase matching, a microfiber with a large $D$ and thus a higher effective index is preferred since the cavity mode has

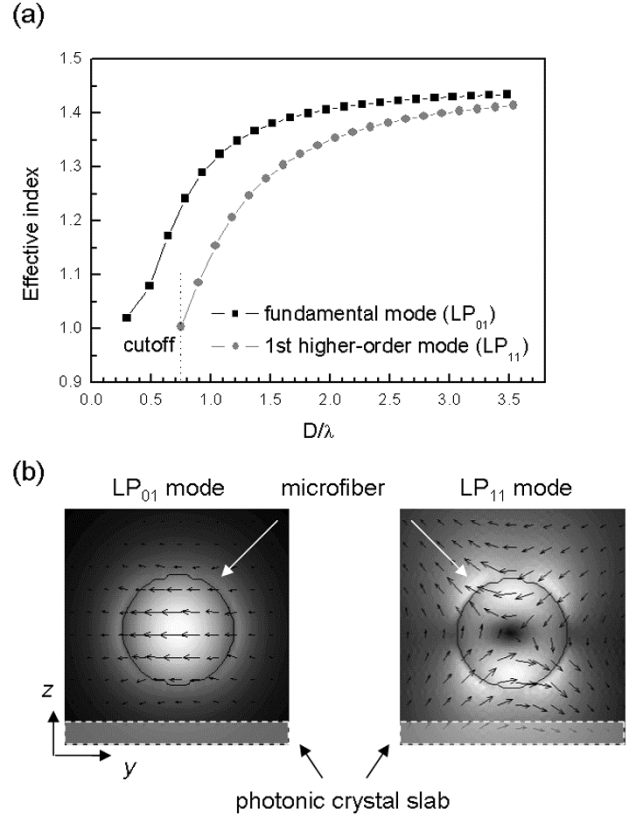

Fig. 2 (a) Dispersion curves of fundamental and first-higher-order modes propagating along the fiber with a diameter of D. (b) Cross-sectional view of electrical field of the two fiber modes which have nonzero mode overlap with the cavity mode. The arrows denote the polarization directions.

relatively high effective index. On the other hand, the mode overlap is enhanced for a microfiber with a smaller $D$ where the mode is weakly guided and the evanescent field extends farther to the photonic crystal cavity. These arguments suggest that there exists an optimal fiber diameter where the coupling strength is maximized.

The photonic crystal cavity structure used in this study is shown in Fig. 3 (a). It has been modified from a 5-lattice-long linear cavity[11] so that the fiber coupling strength($1/Q_{LP01}$) is enhanced while the radiation into air($1/Q_{air}$) is suppressed.[8] The slab thickness was 0.6 $a$, where $a$ is a lattice constant. The resonant mode with the lowest frequency was found at $a/\lambda=0.328$ with an intrinsic $Q$ factor of 87,000. Here we assumed that $a=500$ nm and the resonance wavelength was found to be 1524 nm. The $E_y$ field profile and its Fourier components are shown in Fig. 3 (a) and (b), respectively. The peak in the Fourier space was positioned at $k_x \sim 0.5$ ($2\pi/a$) which corresponds to the effective refractive index of $k_x/(2\pi/\lambda) \sim 1.5$. The white dotted circle is the light cone, which shows the Fourier components of the cavity mode being coupled to the air and contributing to $1/Q_{air}$. The Fourier component of the $LP_{01}$ fiber mode for $D/\lambda=0.7$ is indicated by the black ellipse. Its narrow width and finite height are from infinite extent of the

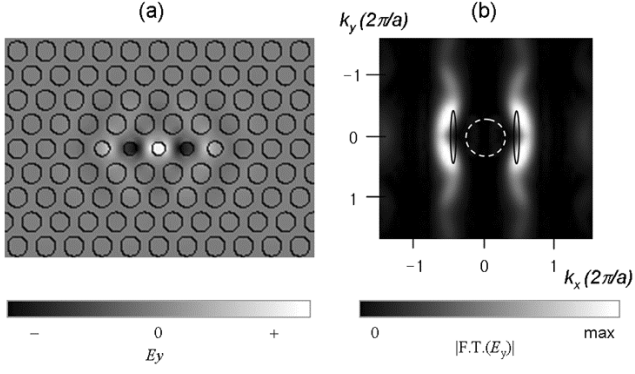

Fig. 3 (a) Photonic crystal cavity structure specially designed for efficient fiber coupling and the Ey field profile of the lowest-frequency mode. The hole sizes were r1=0.35 a for the three middle holes, r2=0.27 a for the two side holes, and r3=0.25 a for others. (b) Fourier-transformed plot of (a). The white circle and the black ellipses denote light cone and the Fourier components of a fiber mode with neff=1.2, respectively.

fiber mode along the fiber length and Gaussian-like confinement along the transverse direction, respectively. The Fourier components overlapped with the black ellipse participate in the coupling with the fiber mode. The large intensity contrast between the two areas in the white circle and the black ellipse indicates that the cavity was effectively designed to suppress the coupling to air while satisfying reasonable (although, still not excellent) phase matching condition with the fiber mode. The Fourier component of the $LP_{11}$ mode which has smaller effective index than the $LP_{01}$ mode should be located between the white circle and the black ellipse in the Fourier space, and the phase matching is not as good as that of the $LP_{01}$ mode.

### III. COMPUTATIONAL SCHEME

To understand the optical coupling from a photonic crystal light source to an optical fiber, the 3-dimensional FDTD method was used. The computational structure is shown in Fig. 4 (a), which included the photonic crystal slab with the cavity in its center and the curved fiber taper with the curvature radius of 50 μm. The computation grid size was one tenth of the lattice constant. The two main parameters we varied in each simulation were the fiber diameter ($D$) and the air gap ($d_{gap}$) between the cavity and the fiber. The cavity mode was excited by the dipole source with the center frequency of 0.328 located inside the cavity.

The total $Q$ of the system was obtained by measuring the decay rate of the cavity mode, and the optical power flow was also monitored at slab edges, fiber facets and air boundaries, to calculate each coupling strength ($1/Q$) in Eq. (2). The power flow integrated over the fiber facet contains the both optical powers in $LP_{01}$ and $LP_{11}$ modes. To selectively measure the power in $LP_{01}$ mode, we used the parity relations of the field components of the LP01 and the LP11 modes, which are observed in Fig. 2 (b). The two modes have the opposite parities in their electrical and magnetic fields respect to the center plane.

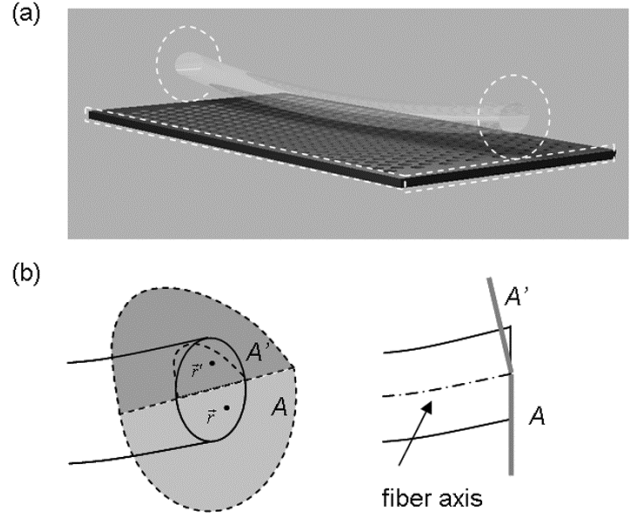

Fig. 4 (a) The computational structure used in FDTD simulation. The pointing vectors were integrated over the dotted rectangles and circles to calculate the powers coupled to the slab, and to the fiber, respectively. (b) Detailed description of the integrated regions used for the calculation of the power portions coupled to $LP_{01}$ and $LP_{11}$ modes. A' is a symmetric plane of A respect to the center plane containing the fiber axis.

For example, the $LP_{01}$ mode has even parity in $E_y$ component while the $LP_{11}$ mode has odd parity, and thus $E_y$ component of $LP_{01}$ mode could be extracted by taking only the symmetric (even) component from the total $E_y$ field as shown in the following Eq. (3). Once all the fields are retrieved to construct the LP01 mode, the optical powers in the two modes, $P_{01}$ and $P_{11}$ can be calculated by integrating the Poynting vector over the fiber cross-section.

$$P_{01} = 2\int_A S_{x01}(\vec{r})da = 2\int_A \left(E_{y01}H_{z01} - H_{y01}E_{z01}\right)da$$

$$P_{11} = \int_{A+A'} S_x(\vec{r})da - P_{01}$$

$$\text{where}\quad E_{y01}(\vec{r}) = \tfrac{1}{2}\left(E_y(\vec{r}) + E_y(\vec{r}')\right) \qquad (3)$$

$$E_{z01}(\vec{r}) = \tfrac{1}{2}\left(E_z(\vec{r}) - E_z(\vec{r}')\right)$$

$$H_{y01}(\vec{r}) = \tfrac{1}{2}\left(H_y(\vec{r}) - H_y(\vec{r}')\right)$$

$$H_{z01}(\vec{r}) = \tfrac{1}{2}\left(H_z(\vec{r}) + H_z(\vec{r}')\right)$$

where $S_x$ and $S_{x01}$ are x-components of Poynting vector of the total field and the $LP_{01}$ mode field, respectively. *A'* and *A* are the upper and the lower halves of the integration area divided by the center plane containing the fiber axis, and they make the same angle with the fiber axis so that $\vec{r}'$ on *A'* becomes the mirror point of $\vec{r}$ on *A* respect to the center plane (See Fig. 4 (b)). The radius of the integral region (*A*, *A'*) was 1.5 μm larger than the fiber radius to take evanescent field of the fiber mode into account.

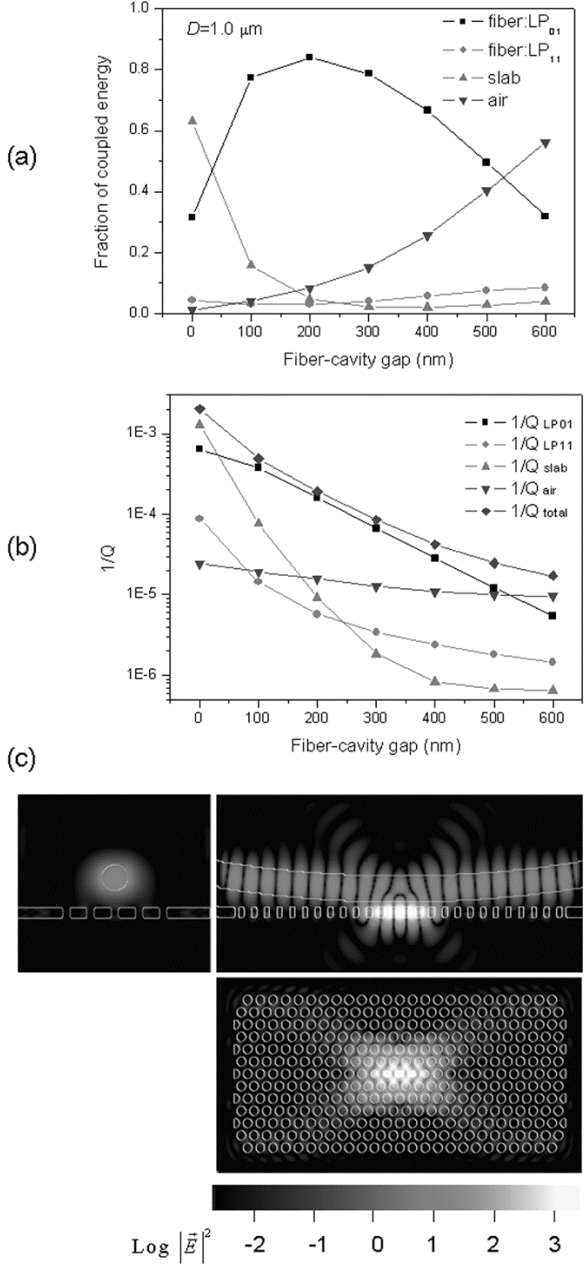

Fig. 3 (a) Fraction of output powers coupled to each section and (b) their coupling strength (inverse of $Q$) plotted as a function of the air gap for the fiber diameter of 1.0 μm. (c) Cross-sectional views of optical intensity profiles at y-z(left), x-z(top) and x-y(botton) planes for the gap of 200 nm. The cavity mode was excited by a dipole source located inside the cavity. It shows significant output coupling into the fundamental mode of fiber and relatively weak radiation into the air and the slab.

## IV. RESULTS AND ANALYSES

First, for the fiber diameter fixed at 1.0 μm, we repeated the simulation to calculate the power flow into each section while varying the fiber-cavity gap ($d_{gap}$), and summarized the results in Fig. 5 (a). As the gap decreased, the more cavity energy tended to couple to the fiber instead of radiating into the air, and the fiber coupling efficiency reached the maximum at $d_{gap}$~200 nm. At the small air gap of $d_{gap}$ < 200 nm, the cavity energy started to leak through the slab and the relative portion of the fiber coupling decreased. The propagating wave along the slab was found in a TM-like mode which originates from the TE-like cavity mode. Note that this TM-like mode does not have a complete bandgap and is allowed to propagate. The TE-to-TM mode conversion occurred when the symmetry of the cavity structure along $z$-axis is broken[12] due to the presence of the fiber near the cavity.

From the data of Fig. 5 (a) and $Q_{tot}$, we calculated the coupling strength ($1/Q$) for each component, and plotted in Fig. 5 (b). Air coupling strength ($Q_{air}$) only slightly changed from its intrinsic value of 87,000 over the variation of $d_{gap}$. The slab coupling strength ($1/Q_{slab}$) was negligible at a large gap, but increased most abruptly as the gap decreased. The coupling to $LP_{01}$ mode ($1/Q_{LP01}$) dominated over other coupling components in the range of $d_{gap}$ of 50~500 nm, and it exhibited exponential decay with $d_{gap}$ reflecting the fact that the evanescent field of the cavity mode decays exponentially from the slab surface. In the fiber with the diameter of 1 μm below the mode cutoff, the $LP_{11}$ mode is not allowed to propagate, and it quickly radiates into air. However, the distance from the cavity to the detection point was not long enough for the photons in $LP_{11}$ to be fully radiated into air, and the non-negligible fraction of the $LP_{11}$ photons still remained in the fiber. This contribution is represented by $1/Q_{LP11}$ in the plot. $Q_{tot}$ which was originally 87,000 decreased down to 492 at $d_{gap}$=0. $Q_{tot}$ for $d_{gap}$=200 nm is 5,200, which shows more than 10 times reduction of $Q$. However, this value is still good for certain microcavity applications.

The snap shot of optical intensity profiles of the coupled system at different cross-sections are depicted in Fig. 5 (c) for $d_{gap}$=200 nm. It is clearly shown that the most of the output power from the cavity is directed into the fiber. The Gaussian-like field profile in the fiber cross-section indicates that the fiber output is mainly in the $LP_{01}$ mode. The weak propagating wave along the slab is the TM-like mode as discussed above. The period of intensity lobes of the fiber mode in $x$-$z$ plane appears to be slightly larger than that of the cavity mode, which visually shows their imperfect phase matching as expected from the Fourier space analyses in Fig. 3 (b).

Next, we investigated the coupling characteristics of fibers with various diameters, $D$. In Fig. 6 (a), the fiber coupling efficiency, $\eta$, was plotted as a function of $d_{gap}$ for a fiber diameter varying from 0.6 to 1.4 μm. In all cases, the coupling peak appeared at $d_{gap}$=200 nm, and the peak value reached the maximum at $D$=1.0 μm. In the same way used above, $1/Q$ for $LP_{01}$ and $LP_{11}$ mode coupling was calculated and plotted in Fig. 6 (b) and (c). $LP_{01}$ mode coupling strength ($1/Q_{LP01}$) always showed the exponential decay as a function of $d_{gap}$ with its absolute magnitude depending on $D$. The coupling strength was maximized at $D$~1.2 um where the $LP_{01}$ mode is relatively well phase matched with the cavity mode while still have strong enough evanescent field out of the fiber. The coupling strength

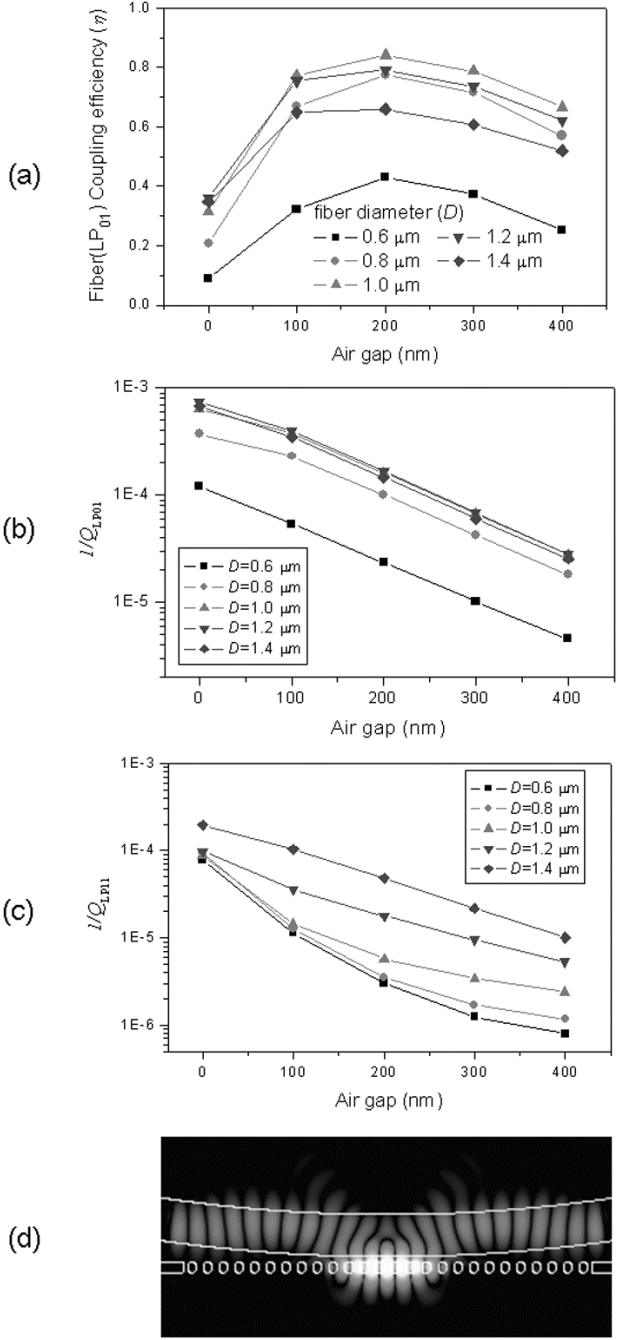

Fig. 4 (a) $LP_{01}$ mode coupling efficiencies, and coupling strengths of (b) $LP_{01}$ mode and (c) $LP_{11}$ mode as functions of the air gap for different fiber diameters. (d) Optical intensity profiles in x-z plane for the fiber diameter of 1.4 mm and the air gap of 200 nm, showing the significant coupling to the $LP_{11}$ fiber mode as well as the $LP_{01}$ mode.

($1/Q_{LP11}$) for the $LP_{11}$ mode increased monotonically with the fiber diameter, which indicates that the $LP_{11}$ mode coupling is limited mainly by the poor phase mismatching rather than the amplitude of the evanescent field. Its exponential dependence on $d_{gap}$ was clear when the $LP_{11}$ mode became truly guided mode above the $LP_{11}$ mode cutoff ($D>1.1$ μm). Other couplings into air and slab did not change significantly with the fiber diameter. Note that the total fiber coupling including $LP_{01}$ and $LP_{11}$ mode coupling increases continuously with the fiber diameter, and if one does not filter out the $LP_{11}$ mode contribution from the total fiber output, it may lead to a wrong conclusion. One example for the case of a large fiber diameter ($D=1.4$ μm) is depicted in Fig. 6 (d). It shows a distorted mode pattern in the fiber due to the modal interference between the $LP_{01}$ and the $LP_{11}$ modes. The coupling efficiency for the $LP_{01}$ mode was limited to 66% although the total fiber coupling efficiency was as large as 88%.

In the previous papers[8,9], the coupling efficiency has been reported without considering the contribution of the $LP_{11}$ mode coupling. In this work, the use of the optimized photonic crystal cavity and fiber diameter resulted in the higher single mode coupling efficiency together with the higher total $Q$ boosted by factor of 3. There are several ways to further improve the coupling efficiency. One of them is to decrease the effective index of the cavity mode for better phase matching with the fiber mode. It can be done by using larger air holes for $r_3$ in Fig. 3 (a) or a thinner slab. The possible drawback is that the intrinsic $Q$ of the cavity mode might be affected as the mode frequency approaches the upper bandedge. Another way is to use a $LP_{01}$-$LP_{11}$ mode coupler at each end of the fiber taper to couple the $LP_{11}$ mode into $LP_{01}$ mode so that all the powers in the $LP_{01}$ and $LP_{11}$ modes can contribute to the single mode output.[13] In ideal case, the coupling efficiency may increase up to ~ 91% for $D=1.8$ um. Or, one can increase the fiber coupling strength simply by using a longer photonic crystal cavity.

## V. CONCLUSIONS

The evanescent coupling between a modified photonic crystal cavity and a curved microfiber has been studied in detail. Two main parameters of the fiber diameter and the air gap between the cavity and the microfiber have been investigated to maximize the cavity-to-fiber coupling efficiency. The use of a microfiber with a larger diameter resulted in a better phase matching between the cavity and the fiber modes and enhanced the coupling strength. However, for a fiber diameter larger than the higher-order mode cutoff (D > 0.75 λ), the coupling to the higher-order fiber mode became nonnegligible, which reduced the relative coupling into the fundamental mode, and the optimal fiber diameter has been found to be 1.0 μm. To increase the mode overlap and the coupling efficiency, the air gap was decreased until the cavity mode started to couple with a slab propagating mode. The coupling efficiency reached the maximum of 84% for the air gap of 200 nm, and the total Q factor of the coupled system was 5,200.

The fiber coupling can also be used to inject the light into photonic crystal cavity through the fiber as well as to collect the output, which is very useful in case of resonant filters or nonlinear switches. In those cases, most of the analyses and discussions on this study are still valid, and the simulation results presented here can be directly used to build the coupled mode equations describing the device responses[14].